\documentclass[aps,prd,twocolumn,nofootinbib,superscriptaddress]{revtex4-2}

\usepackage{amsmath,amssymb,mathrsfs,bm}
\usepackage{booktabs}
\usepackage[colorlinks=true,linkcolor=blue,citecolor=blue,urlcolor=blue]{hyperref}
\usepackage{orcidlink}

\newcommand{\CP}{\mathbb{C}P}
\newcommand{\Tr}{\operatorname{Tr}}
\newcommand{\ind}{\operatorname{index}}
\newcommand{\Spin}{\operatorname{Spin}}
\newcommand{\Spinc}{\mathrm{Spin}^{c}}
\newcommand{\cO}{\mathcal O}

\newcommand{\slD}{\not\!D}
\begin{document}

\title{Three Chiral Families from a Monopole Product over \texorpdfstring{$S^2\times\mathbb{C}P^2$}{S2 x CP2}}

\author{Edward J. Shaya\ \orcidlink{0000-0002-3234-8699}}
\email{eshaya2@gmail.com}
\affiliation{Department of Astronomy, University of Maryland, College Park, USA}

\date{\today}

\begin{abstract}
We study the chiral zero-mode spectrum of the compact six-manifold
$X_6=S^2\times \CP^2$ in a ten-dimensional Kaluza-Klein setting, allowing
for torsionful geometric deformations that preserve ellipticity.  The paper is organized to separate known ingredients from the new
construction.  We first review the established facts: the Witten obstruction to
chirality in simple Kaluza-Klein reductions, the use of background fluxes and
index theory to evade it, and the monopole Dirac index on $S^2$.
Dolan and Nash established the $\mathrm{Spin}^c$ index mechanism on complex projective spaces and showed that $\mathbb{C}P^2$ can supply a one-family chiral building block with Standard-Model-like quantum numbers. The step taken here is to place that known $\mathbb{C}P^2$ sector on the six-dimensional product $S^2\times\mathbb{C}P^2$ and use an independently quantized monopole sector on $S^2$ to control the family multiplicity.

Unlike Calabi--Yau model-building constructions, where the family
number is typically tied to vector-bundle and quotient data, this homogeneous product gives the three-family count through the factorized Dirac-index formula:
$
       N_{\rm fam} = \ind \slD_{S^2\times \CP^2}
        =\bigl(\ind \slD_{S^2,m}\bigr)
         \,\bigl(\ind \slD_{\CP^2,n}\bigr)
        =m\,\frac{n^2-1}{8}.
$
The canonical complex $\Spinc$ structure on $\CP^2$ has $n=3$ and unit index;
using the Lichnerowicz-Dolbeault form of the $\Spinc$ Dirac operator, we show
that this unit index is saturated by exactly one chiral zero mode and no
opposite-chirality vectorlike partner.  Therefore two flux sectors produce exactly three chiral
internal zero modes: $(m,n)=(3,3)$, where the monopole on $S^2$
supplies the multiplicity, and $(m,n)=(1,5)$, where the higher
$\Spinc$ twist on $\CP^2$ does.  Saturation---the absence of
vectorlike partners, so that the index equals the kernel
dimension---is proved for all odd $n\geq3$ by Kodaira vanishing on
the twisted spinor bundle.

For a ten-dimensional Weyl fermion valued in a visible $\mathbf{16}$ of
$\Spin(10)_G$, the Lorentz branching
$\Spin(1,9)\supset\Spin(1,3)\times\Spin(6)$ shows that these internal
zero modes give three left-handed four-dimensional $\mathbf{16}_G$ multiplets,
up to the conventional choice of ten-dimensional chirality.  We formulate the ten-dimensional action,
state the flux quantization conditions, write the torsionful Dirac operator,
and estimate the leading Kaluza-Klein scale.  The analysis is restricted to the
spectral and index-theoretic problem; moduli stabilization, Yukawa textures, and
other phenomenological dynamics are not derived here.
\end{abstract}

\maketitle

\section{Introduction}

A long-standing obstacle for Kaluza-Klein unification is the appearance of
vectorlike rather than chiral fermion spectra after compactification
\cite{Witten1981}.  Background gauge fields and nontrivial spin or $\Spinc$
structures provide a way around this obstruction: the net number of chiral
four-dimensional fermions is controlled by the index of an internal Dirac
operator rather than by local curvature alone \cite{AtiyahSinger1968,
RandjbarDaemi1983,Dolan1985,DolanNashSpinC2002,DolanNash2002}.

This paper studies one sharply defined problem.  Let the ten-dimensional
manifold be
\begin{equation}
        M_{10}=M_4\times X_6,
        \qquad X_6=S^2\times \CP^2 .
\end{equation}
We ask whether the internal Dirac operator on $X_6$ can supply exactly three
chiral four-dimensional zero modes in a controlled topological sector.

For clarity about priority, Sec.~\ref{sec:history} is a review section.  It
collects the Dolan-Nash results, standard index-theorem facts, the monopole
Dirac index on $S^2$, and the generic $\Spin(10)$ anomaly-free family.  No
novelty is claimed for that material.  The new construction begins in
Sec.~\ref{sec:newsetup}.  Its central claim is narrower than a derivation of the
full Standard Model: the known unit-index $\CP^2$ $\Spinc$ sector can be used as
a one-family building block, and the monopole number on the additional $S^2$
factor then multiplies this block.  For the canonical complex $\Spinc$ structure
on $\CP^2$ and monopole number three on $S^2$, the factorized internal Dirac
index is exactly three.

Three-family chiral spectra are known in several compactification
frameworks, most notably in heterotic Calabi--Yau constructions with
nontrivial vector bundles, quotients, and Wilson lines
\cite{Candelas1985,Anderson2012}.  The present construction is not
intended as a uniqueness theorem over all compactifications.  Its novelty is different: to our knowledge,
$S^2\times\CP^2$ is the only currently known homogeneous
six-dimensional product geometry for which a three-family chiral
spectrum follows from such a direct factorization of Dirac indices, with one factor
supplying an $S^2$ monopole multiplicity and the other supplying the
minimal $\Spinc$ chiral block on $\CP^2$.  In this sense the
construction isolates the family number in an elementary geometric
formula rather than in a scan over Calabi--Yau metrics, bundles, or
Wilson-line data.

This separation is important.  The generation count is topological and robust
under smooth deformations of the connection, including torsionful deformations
that preserve ellipticity.  By contrast, stabilization of the internal radii, the Higgs sector, Yukawa
textures, and mass generation are dynamical questions.  They are not needed for
the index calculation and are therefore not claimed in this paper.

\section{Known results and historical background}
\label{sec:history}

We begin by reviewing what has been uncovered thus far about this manifold.
More precisely, we review the established facts about its two factors,
$S^2$ and $\CP^2$, and about the standard gauge-theory embedding used later.
This section fixes notation and assigns credit for the ingredients on which the
new product construction is built.

\subsection{Chirality, flux, and the index theorem}

Witten showed that simple Kaluza-Klein compactifications face a severe
chirality obstruction: for many compact internal manifolds, the low-energy
fermion spectrum is vectorlike rather than chiral \cite{Witten1981}.  A standard
way around this obstruction is to include topologically nontrivial background
gauge fields.  The number of net chiral zero modes is then governed by the
Atiyah-Singer index theorem \cite{AtiyahSinger1968}.  This mechanism underlies
many coset-space and flux compactification constructions
\cite{RandjbarDaemi1983,Forgacs1980,Kapetanakis1992}.

\subsection{The $\CP^2$ factor and the Dolan-Nash result}

The facts in this subsection are standard, and their use in chiral model
building on complex projective spaces is due in particular to Dolan and Nash
\cite{DolanNashSpinC2002,DolanNash2002}.  The cohomology generator of
$H^2(\CP^2,\mathbb Z)$ will be denoted by $v$, normalized by
\begin{equation}
        \int_{\CP^2}v^2=1 .
\end{equation}
The characteristic classes are
\begin{equation}
        c_1(T\CP^2)=3v,
        \qquad c_2(T\CP^2)=3v^2,
\end{equation}
so
\begin{equation}
        p_1(T\CP^2)=c_1^2-2c_2=3v^2 .
\end{equation}
The second Stiefel-Whitney class is
\begin{equation}
        w_2(T\CP^2)=c_1(T\CP^2)\bmod 2=v\bmod 2 .
\end{equation}
Thus $\CP^2$ is not spin.  It does, however, admit $\Spinc$ structures.  A
$\Spinc$ determinant line $L_n\rightarrow \CP^2$ with
\begin{equation}
        c_1(L_n)=n v
\end{equation}
is admissible precisely when
\begin{equation}
        n\equiv 1 \pmod 2 .
\end{equation}
The canonical complex $\Spinc$ structure associated with the complex structure
has determinant line $K^{-1}=\cO(3)$, hence $n=3$.

The $\Spinc$ Dirac index on $\CP^2$ is
\begin{equation}
        \ind\slD_{\CP^2,n}
        =\int_{\CP^2} e^{c_1(L_n)/2}\,\hat A(T\CP^2).
\end{equation}
Using
\begin{equation}
        \hat A(T\CP^2)=1-\frac{p_1}{24}=1-\frac{3v^2}{24},
\end{equation}
and
\begin{equation}
        e^{nv/2}=1+\frac{n}{2}v+\frac{n^2}{8}v^2,
\end{equation}
one obtains
\begin{equation}
        \ind\slD_{\CP^2,n}
        =\frac{n^2-1}{8} .
        \label{eq:knowncp2index}
\end{equation}
This is an integer for odd $n$.  For the canonical complex $\Spinc$ structure,
$n=3$, Eq.~\eqref{eq:knowncp2index} gives
\begin{equation}
        \ind\slD_{\CP^2,3}=1 .
\end{equation}
Dolan and Nash showed that such $\Spinc$ sectors on complex projective spaces
can be used to obtain chiral zero modes with Standard-Model-like quantum
numbers, including a right-handed neutrino \cite{DolanNashSpinC2002,
DolanNash2002}.  In the present paper this is not treated as a new result; it
is the known one-family $\CP^2$ building block.

\subsection{The $S^2$ monopole Dirac index}

The $S^2$ ingredient is also standard.  Let $u$ generate
$H^2(S^2,\mathbb Z)$, normalized by
\begin{equation}
        \int_{S^2}u=1 .
\end{equation}
A monopole line bundle $E_m\rightarrow S^2$ has
\begin{equation}
        c_1(E_m)=m u,
        \qquad m\in \mathbb Z,
\end{equation}
or equivalently
\begin{equation}
        \frac{1}{2\pi}\int_{S^2}F_{S^2}=m .
\end{equation}
The spin Dirac index on $S^2$ twisted by $E_m$ is
\begin{equation}
        \ind\slD_{S^2,m}=
        \int_{S^2}\mathrm{ch}(E_m)\,\hat A(TS^2)=m,
        \label{eq:knowns2index}
\end{equation}
since $\hat A(TS^2)=1$ in degree two.

For a round sphere of radius $a_1$, the corresponding monopole Dirac spectrum has
zero-mode degeneracy $|m|$ and a massive tower whose leading scale is set by
$a_1^{-1}$.  With the common normalization
\begin{equation}
        \lambda^2_{k,S^2}=\frac{k(k+|m|)}{a_1^2},
        \qquad k=0,1,2,\ldots,
        \label{eq:knowns2spectrum}
\end{equation}
with degeneracy
\begin{equation}
        d_k=|m|+2k,
\end{equation}
the first nonzero level is
\begin{equation}
        M_{1,S^2}=\frac{\sqrt{|m|+1}}{a_1} .
\end{equation}
The sign of $m$ selects the internal chirality of the zero modes.

\subsection{Torsion and index stability}

The general torsionful Dirac operator on a Riemannian manifold may be written
schematically as
\begin{equation}
\slD_T=\Gamma^a\left(\partial_a+\frac{1}{4}\omega^{\rm LC}_{abc}\Gamma^{bc}
+\frac{1}{4}K_{abc}\Gamma^{bc}+A_a\right),
\end{equation}
where $K_{abc}$ is the contorsion tensor.  Smooth torsion deformations that
preserve ellipticity can move nonzero eigenvalues and alter wavefunctions, but
they do not change the topological index.  This is a standard consequence of
the homotopy invariance of the index, with local-index refinements available in
the presence of torsion \cite{Bismut1989,LawsonMichelsohn1989,Friedrich2000}.
The present paper therefore does not use torsion to create the generation
number; torsion is included only to make explicit that smooth torsionful
geometric deformations do not change the index.

\subsection{The standard $\Spin(10)$ family and anomaly cancellation}

The use of a $\mathbf{16}$ of a visible $\Spin(10)_G$ gauge group to package one
Standard Model family plus a right-handed neutrino is also standard.  Under
\begin{equation}
        \Spin(10)_G\supset SU(3)_C\times SU(2)_L\times U(1)_Y,
\end{equation}
one left-handed family decomposes as
\begin{equation}
\begin{split}
\mathbf{16} \rightarrow&\ (3,2)_{1/6}
\oplus (\bar 3,1)_{-2/3}
\oplus (\bar 3,1)_{1/3} \\
&\oplus (1,2)_{-1/2}
\oplus (1,1)_1
\oplus (1,1)_0 .
\end{split}
\label{eq:knownspin10decomp}
\end{equation}
The final singlet is the left-handed conjugate of a right-handed neutrino.

The corresponding four-dimensional gauge and mixed anomalies cancel family by
family.  In left-handed notation,
\begin{align}
A_{331}&=2T(3)Y_Q+T(\bar 3)Y_{u^c}+T(\bar 3)Y_{d^c}=0,\\
A_{221}&=3T(2)Y_Q+T(2)Y_L=0,
\end{align}
where $T(3)=T(\bar 3)=T(2)=1/2$.  

The mixed gravitational-$U(1)_Y$ anomaly is
\begin{equation}
        \sum_i d_iY_i=0,
\end{equation}
and the cubic hypercharge anomaly is
\begin{equation}
        \sum_i d_iY_i^3=0 .
\end{equation}
Expanded over the fields in Eq.~\eqref{eq:knownspin10decomp}, these are the
usual one-family Standard Model cancellations.  The global $SU(2)$ anomaly is absent because
one family contains four left-handed $SU(2)$ doublets, counting the three colors
of $Q_L$ plus the lepton doublet $L_L$.  Thus three copies of such a family are
also anomaly-free.

\section{New construction: ten-dimensional product compactification}
\label{sec:newsetup}

We now use the reviewed ingredients to define the product compactification.  The
Lorentzian ten-dimensional tangent group is $\Spin(1,9)$, which must be kept
distinct from the visible gauge group $\Spin(10)_G$.  Fermions are sections of a
tensor product of the ten-dimensional spinor bundle, the internal $\Spinc$
bundle, and a representation of the visible gauge group.

A minimal effective Einstein-Cartan-Yang-Mills action sufficient for the
zero-mode analysis is
\begin{widetext}
\begin{equation}
S_{10}=\int_{M_{10}} d^{10}x\,e\left[
 \frac{1}{2\kappa_{10}^2}\big(R(\omega)-2\Lambda_{10}\big)
 -\frac{1}{4g_{10}^2}\Tr(F_{MN}F^{MN})
 +\frac{i}{2}\bar\Psi\Gamma^M\overleftrightarrow{D}_M^{(\omega,A)}\Psi
 \right],
\label{eq:10daction}
\end{equation}
\end{widetext}
where $e=\sqrt{-g_{10}}$, $M,N=0,\ldots,9$, and
\begin{equation}
D_M^{(\omega,A)}=\partial_M+\frac{1}{4}\omega_{MAB}\Gamma^{AB}+A_M .
\end{equation}
The spin connection is allowed to have torsion,
\begin{equation}
        \omega_{MAB}=\omega^{\rm LC}_{MAB}+K_{MAB},
\end{equation}
where $K_{MAB}$ is the contorsion tensor.  The index calculation below depends
only on the elliptic internal Dirac operator and its topological twisting data.

The internal space is
\begin{equation}
        X_6=S^2_{a_1}\times \CP^2_{a_2},
\end{equation}
where $a_1$ is the radius of the two-sphere and $a_2$ denotes the curvature
scale of the Fubini-Study metric on $\CP^2$.

\subsection{Splitting of the internal gauge connection}
\label{subsec:connection-splitting}

The symbol $A_a$ in the internal Dirac operator denotes the full connection
acting on the internal part of the fermion bundle.  It is useful to separate it
into visible gauge pieces and topological line-bundle pieces.  For a fermion in
representation $\rho$ of the visible gauge group, we write schematically
\begin{align}
        A_a^{\rm tot}={}& A_{a,C}^{A}\,\rho(T_C^A)
        + A_{a,L}^{i}\,\rho(T_L^i)
        + A_{a,Y}\,\rho(Y)\nonumber\\
        &+ i q_m a^{(m)}_a
        + \frac{i}{2}a^{(n)}_a .
\label{eq:connection-split}
\end{align}
Here $T_C^A$, $T_L^i$, and $Y$ are the generators of
$SU(3)_C$, $SU(2)_L$, and $U(1)_Y$, respectively.  The first three terms are the
visible Standard Model gauge connection, or possible internal background pieces
in the corresponding sectors.  In the minimal product construction used below
we take the color and electroweak internal backgrounds to vanish, so that color
and hypercharge label the four-dimensional multiplets rather than provide the
topological twist.  The two remaining terms are different: $a^{(m)}$ is the
ordinary monopole connection on $S^2$, while $a^{(n)}$ is a connection on the
$\Spinc$ determinant line $L_n\rightarrow \CP^2$.  The factor of $1/2$ in the
last term is essential.  A $\Spinc$ spinor couples to one half of the determinant-line
connection; hence the effective magnetic flux seen by the Dirac operator is
half the integral determinant-line flux.  This topological $\Spinc$ connection
should not be identified with Standard Model hypercharge.

Let $u\in H^2(S^2,\mathbb Z)$ and $v\in H^2(\CP^2,\mathbb Z)$ obey
\begin{equation}
        \int_{S^2}u=1,
        \qquad
        \int_{\CP^1}v=1,
        \qquad
        \int_{\CP^2}v^2=1 .
\end{equation}
The product background uses the $S^2$ monopole line bundle $E_m$ and a
$\Spinc$ determinant line $L_n$ on $\CP^2$,
\begin{equation}
        c_1(E_m)=m u,
        \qquad
        c_1(L_n)=n v .
\label{eq:product-line-classes}
\end{equation}
In flux language,
\begin{equation}
        \frac{1}{2\pi}\int_{S^2} f^{(m)}=m,
        \qquad
        \frac{1}{2\pi}\int_{\CP^1} f^{(n)}=n,
\label{eq:det-line-flux}
\end{equation}
where $f^{(m)}=d a^{(m)}$ and $f^{(n)}=d a^{(n)}$ are ordinary integral
curvatures.  However, because the $\Spinc$ spinor couples to $a^{(n)}/2$, the
flux appearing in the $\CP^2$ Dirac operator is
\begin{equation}
        \frac{1}{2\pi}\int_{\CP^1}\frac{f^{(n)}}{2}=\frac{n}{2} .
\label{eq:half-flux}
\end{equation}
Thus an odd determinant class $n$ produces a half-integer magnetic flux in the
fermion operator.

The reason this half-integrality is required is topological.  For an oriented
manifold $M$, a $\Spinc$ structure with determinant line $L$ exists only if
\begin{equation}
        c_1(L)\equiv w_2(TM) \pmod 2 .
\label{eq:spinc-condition}
\end{equation}
This condition says that the mod-two reduction of the determinant line cancels
the obstruction to lifting the oriented orthonormal frame bundle from $SO(d)$ to
$\Spin(d)$.  Equivalently, on triple overlaps the sign obstruction in the spin
transition functions is compensated by the sign in the square root of the
$U(1)$ transition functions.  This is the precise sense in which the
half-integer $\Spinc$ flux cancels the obstruction to defining the Dirac
operator.

For $\CP^2$, the tangent bundle has
\begin{align}
        c_1(T\CP^2)&=3v,\\
        w_2(T\CP^2)&=c_1(T\CP^2)\bmod 2=v\bmod 2 .
\end{align}
Therefore Eq.~\eqref{eq:spinc-condition} becomes
\begin{equation}
        n v\equiv v \pmod 2,
        \qquad\text{or equivalently}\qquad
        n\in 2\mathbb Z+1 .
\label{eq:n-odd-condition}
\end{equation}
Any odd $n$ cancels the topological obstruction.  The value $n=3$ is special
not because it is the only allowed $\Spinc$ class, but because it is the
canonical complex $\Spinc$ structure on $\CP^2$ and, as reviewed above, has unit
Dirac index.  It is therefore the minimal positive-index $\CP^2$ block used in
the product construction.

On the product background, the ten-dimensional Dirac operator separates as
\begin{equation}
        \slD_{10}=\slD_4\otimes 1+\gamma_5\otimes \slD_{X_6,T},
\end{equation}
up to warp-factor and mixing terms.  The internal torsionful operator is
\begin{equation}
\slD_{X_6,T}=\Gamma^a\left(\partial_a+\frac{1}{4}\omega^{\rm LC}_{abc}\Gamma^{bc}
+\frac{1}{4}K_{abc}\Gamma^{bc}+A_a\right),
\label{eq:torsiondirac}
\end{equation}
where $a,b,c$ are internal indices and $A_a$ is the total connection in
Eq.~\eqref{eq:connection-split}.  The role of torsion here is geometric and
dynamical; the net chiral index is still fixed by the flux and $\Spinc$ classes.

\section{Product index and the three-family sector}
\label{sec:productindex}

The index of the product Dirac operator factorizes.  Using the reviewed factor
indices, Eqs.~\eqref{eq:knowncp2index} and \eqref{eq:knowns2index}, gives
\begin{equation}
\boxed{
\begin{gathered}
        \ind\slD_{S^2\times\CP^2}
        =\ind\slD_{S^2,m}\,\ind\slD_{\CP^2,n}\\
        \ind\slD_{S^2\times\CP^2}=m\,\frac{n^2-1}{8} .
\end{gathered}}
\label{eq:mainindex}
\end{equation}
The point of the product construction is that the $\CP^2$ factor supplies a
known unit-index chiral block, while the $S^2$ monopole number supplies a
controlled replication factor.

For the canonical complex $\Spinc$ structure on $\CP^2$, $n=3$, and hence
\begin{equation}
        \ind\slD_{\CP^2,3}=1 .
\end{equation}
For monopole number $m=3$ on $S^2$, Eq.~\eqref{eq:mainindex} gives
\begin{equation}
\boxed{
        \ind\slD_{S^2\times\CP^2}=3 .
}
\end{equation}
Thus the proposed mechanism is not that $\CP^2$ alone explains three families.
Rather,
\begin{equation}
\begin{array}{c}
        \hbox{one Dolan--Nash-type }\CP^2\hbox{ chiral block}\\[2pt]
        \times\ \hbox{monopole number }m=3
\end{array}
\end{equation}
produces three replicated internal zero modes on the product space.

With the orientation chosen so that $\int_{S^2}u=1$ and
$\int_{\CP^2}v^2=1$, positive $m$ and positive $n=3$ produce positive internal
chirality.  Reversing the monopole charge reverses the sign of the index,
\begin{equation}
        \ind\slD_{X_6}(-m,n)=-\ind\slD_{X_6}(m,n).
\end{equation}
The absolute number of zero modes is at least $|\ind\slD|$.  Equality requires
absence of additional vectorlike zero-mode pairs.  For the present choice this
can be shown analytically, as we now explain.

\subsection{Lichnerowicz proof of saturation for $n=3$}
\label{sec:nopairs}

The remaining issue is whether the index-one $\CP^2$ sector hides additional
opposite-chirality zero modes.  For $n=3$ it does not.  The reason is that
$n=3$ is the canonical complex $\Spinc$ structure on $\CP^2$.  Since the
canonical bundle is
\begin{equation}
        K_{\CP^2}={\cal O}(-3),
        \qquad K_{\CP^2}^{-1}={\cal O}(3),
\end{equation}
the determinant line of the canonical complex $\Spinc$ structure is precisely
$L_3=K_{\CP^2}^{-1}$, so $c_1(L_3)=3v$.  The associated spinor bundle is
\begin{equation}
        {\cal S}_{3}\simeq \Lambda^{0,*}T^*\CP^2,\,\, {\cal S}_{3}^{+}=\Lambda^{0,0}\oplus \Lambda^{0,2},\,\, {\cal S}_{3}^{-}=\Lambda^{0,1}.
\end{equation}
With the Chern connection on $K^{-1}_{\CP^2}$, the $\Spinc$ Dirac operator is
identified with the Dolbeault-Dirac operator,
\begin{equation}
        \slD_{\CP^2,3}=\sqrt{2}\,
        \left(\bar\partial+\bar\partial^\dagger\right).
\label{eq:dolbeaultdirac}
\end{equation}
Equivalently, the $\Spinc$ Lichnerowicz formula~\cite{LawsonMichelsohn1989,Friedrich2000,DolanNashSpinC2002}
\begin{equation}
        \slD_{L}^{2}=\nabla_{L}^{*}\nabla_{L}
        +\frac{{\cal R}}{4}+\frac{i}{2}c(F_L)
\label{eq:spinclichnerowicz}
\end{equation}
reduces, for the canonical complex $\Spinc$ connection, to the Dolbeault
Weitzenbock identity
\begin{equation}
        \slD_{\CP^2,3}^{2}=2\Delta_{\bar\partial}.
\end{equation}
Here $F_L$ is the curvature of the determinant-line connection and $c(F_L)$
denotes Clifford multiplication.  Therefore the zero modes are exactly the
Dolbeault harmonic forms,
\begin{equation}
        \ker \slD_{\CP^2,3}\simeq
        H^{0,0}(\CP^2)\oplus H^{0,1}(\CP^2)\oplus H^{0,2}(\CP^2),
\end{equation}
with chirality given by the parity of the form degree.  The Hodge numbers of
$\CP^2$ are
\begin{equation}
        h^{0,0}=1,
        \qquad h^{0,1}=0,
        \qquad h^{0,2}=0.
\end{equation}
Thus
\begin{equation}
        \dim\ker \slD_{\CP^2,3}^{+}=1,
        \,\,
        \dim\ker \slD_{\CP^2,3}^{-}=0,
\label{eq:cp2saturated}
\end{equation}
and the index is saturated by a single positive-chirality zero mode.  There are
no hidden vectorlike pairs in the $n=3$ $\CP^2$ sector.

The same conclusion holds on the $S^2$ factor for positive monopole number:
the standard monopole Dirac theorem gives
\begin{equation}
        \dim\ker \slD_{S^2,m}^{+}=m,
        \,\,
        \dim\ker \slD_{S^2,m}^{-}=0,
        \,\, m>0 .
\end{equation}
For the product operator,
\begin{equation}
        \slD_{X_6}^{2}=\slD_{S^2}^{2}\otimes 1+1\otimes
        \slD_{\CP^2}^{2},
\end{equation}
so a product zero mode must be a zero mode of both factor operators.  Combining
the preceding equations gives
\begin{align}
        \dim\ker \slD_{S^2\times\CP^2}^{+}&=m,\\
        \dim\ker \slD_{S^2\times\CP^2}^{-}&=0,
        \qquad (m>0,n=3).
\end{align}
For $m=3$, the product has exactly three chiral zero modes, not merely index
three.

\subsection{Saturation for all odd $n\geq3$}
\label{subsec:general-saturation}

The Lichnerowicz--Dolbeault proof above exploits the identification
$L_3 = K^{-1}$, which is special to the canonical class $n=3$.  A
natural question is whether the same saturation---index equals kernel
dimension, with no vectorlike pairs---holds for non-canonical odd
classes $n\geq5$.  It does, by a short cohomological argument that
also answers the complementary question of whether the positive-chirality
kernel could be \emph{larger} than the index.

For general odd $n\geq3$, the $\Spinc$ spinor bundle on $\CP^2$ can
be written
\begin{equation}
        S_n = S_3\otimes\cO(k),\qquad k=\frac{n-3}{2}\geq0,
\label{eq:Sn-twist}
\end{equation}
where $S_3$ is the canonical $\Spinc$ bundle (the ordinary Dolbeault
complex) and $\cO(k)$ is the $k$-th power of the hyperplane bundle.
The chirality decomposition is
\begin{equation}
        S_n^+=\cO(k)\oplus\Omega^{0,2}(k),
        \qquad
        S_n^-=\Omega^{0,1}(k),
\label{eq:Sn-grading}
\end{equation}
so the harmonic kernels are Dolbeault cohomology groups:
\begin{equation}
\begin{split}
\ker\slD\big|_{S_n^+}&=H^{0,0}\bigl(\cO(k)\bigr) \oplus H^{0,2}\bigl(\cO(k)\bigr),\\
\ker\slD\big|_{S_n^-}&=H^{0,1}\bigl(\cO(k)\bigr).
\end{split}
\label{eq:kernel-cohomology}
\end{equation}
For $k\geq0$ on $\CP^2$:
\begin{enumerate}
\item $H^{0,1}\bigl(\cO(k)\bigr)=0$: for $k>0$ this is Kodaira
        vanishing ($\cO(k)$ is ample); for $k=0$ it follows from
        $h^{0,1}(\CP^2)=0$.
\item $H^{0,2}\bigl(\cO(k)\bigr)\cong H^0\bigl(\cO(-k-3)\bigr)^\vee
        =0$ by Serre duality, since $-k-3<0$.
\item $H^{0,0}\bigl(\cO(k)\bigr)=H^0\bigl(\cO(k)\bigr)
        =\binom{k+2}{2}
        =\frac{(n+1)(n-1)}{8}=\frac{n^2-1}{8}$.
\end{enumerate}
Therefore, for every odd $n\geq3$,
\begin{equation}
        \ker^-=0,\qquad
        \dim\ker^+=\frac{n^2-1}{8}=\ind\slD_n.
\label{eq:general-saturation}
\end{equation}
The index is saturated: there are no vectorlike partners, and the
positive-chirality kernel cannot exceed the index because the only
contributing cohomology group is $H^{0,0}$, whose dimension is fixed
by the degree of the line bundle.

On the product $S^2\times\CP^2$, the six-dimensional Dirac kernel
factorizes into tensor products of like-chirality factor kernels
(both factors have $\ker^-=0$), so
$\dim\ker_{6D}=m\times(n^2-1)/8$ with no additional non-product
zero modes.  The product index formula
$N_{\rm fam}=m(n^2-1)/8$ therefore gives the \emph{exact} number of
chiral zero modes for all admissible $(m,n)$, not merely a lower
bound.

In particular, this establishes a second saturated three-family
sector, $(m,n)=(1,5)$, degenerate with $(3,3)$ in family number; its
kernel is worked out explicitly in the next subsection.

\subsection{Explicit saturation study of the $(m,n)=(1,5)$ branch}
\label{subsec:one-five}

The second three-family solution is sufficiently close to the
preferred $(3,3)$ branch that it is useful to record its kernel
explicitly.  For $n=5$ one has $k=(n-3)/2=1$, so the positive-chirality
$\CP^2$ spinor bundle is
\begin{equation}
        S_{5}^{+}=\cO(1)\oplus\Omega^{0,2}(1),
        \qquad
        S_{5}^{-}=\Omega^{0,1}(1).
\end{equation}
The relevant Dolbeault cohomology groups are
\begin{equation}
        H^{0,1}\bigl(\cO(1)\bigr)=0,
        \qquad
        H^{0,2}\bigl(\cO(1)\bigr)=0,
\end{equation}
and
\begin{equation}
        H^{0,0}\bigl(\cO(1)\bigr)=H^0\bigl(\CP^2,\cO(1)\bigr)
        \cong \mathbb C^3 .
\end{equation}
Thus
\begin{equation}
        \dim\ker\slD_{\CP^2,5}^{+}=3,
        \qquad
        \dim\ker\slD_{\CP^2,5}^{-}=0,
\end{equation}
so the $\CP^2$ index
\begin{equation}
        \frac{5^2-1}{8}=3
\end{equation}
is exactly saturated by three positive-chirality modes.  In homogeneous
coordinates $[Z_0:Z_1:Z_2]$ on $\CP^2$, these modes may be represented,
up to the common spinor factor and normalization, by the linear sections
\begin{equation}
        Z_0,\qquad Z_1,\qquad Z_2
        \quad\in H^0\bigl(\CP^2,\cO(1)\bigr).
\end{equation}
Equivalently, they transform as the fundamental $\mathbf 3$ of the
$SU(3)$ isometry of $\CP^2$.

For $m=1$ the $S^2$ monopole operator contributes exactly one
positive-chirality zero mode and no negative-chirality partner,
\begin{equation}
        \dim\ker\slD_{S^2,1}^{+}=1,
        \qquad
        \dim\ker\slD_{S^2,1}^{-}=0.
\end{equation}
The product kernel is therefore
\begin{equation}
\begin{split}
\ker\slD_{S^2\times\CP^2}^{+}&=\ker\slD_{S^2,1}^{+}\otimes
H^0\bigl(\CP^2,\cO(1)\bigr),\\
\ker\slD_{S^2\times\CP^2}^{-}&=0,
\end{split}
\end{equation}
with
\begin{equation}
        \dim\ker\slD_{S^2\times\CP^2}^{+}=1\times3=3.
\end{equation}
Thus $(m,n)=(1,5)$ is a genuine saturated three-family branch, not an
index-only branch hiding vectorlike pairs.  Its distinction from
$(m,n)=(3,3)$ is geometric rather than numerical: in $(3,3)$ the
family triplet lives on the $S^2$ monopole tower, while in $(1,5)$ it
lives on the $\CP^2$ linear-section tower.  The two branches therefore
produce the same net chiral family count but different internal
wave-function profiles and, consequently, different higher-dimensional
overlap tensors.

\section{Four-dimensional zero-mode spectrum}
\label{sec:spectrum}

\subsection{Ten-dimensional Lorentz spinors and four-dimensional chirality}
\label{subsec:lorentz-branching}

We now spell out a point of notation that is essential for the spectrum.  The
symbol $\mathbf{16}$ can refer to two different objects: a chiral Lorentz spinor
of $\Spin(1,9)$, or the spinor representation of a visible gauge group
$\Spin(10)_G$.  We distinguish them by writing the ten-dimensional Lorentz
spinor as $S_{10}^{\pm}$ and the visible gauge representation as
$\mathbf{16}_G$.

The ten-dimensional tangent-group branching is
\begin{equation}
        \Spin(1,9)\supset \Spin(1,3)\times \Spin(6),
        \qquad \Spin(6)\cong SU(4).
\end{equation}
Let
\begin{equation}
        S_4^L=(2,1),\qquad S_4^R=(1,2)
\end{equation}
denote four-dimensional left- and right-handed Weyl spinors under
$SL(2,\mathbb C)\simeq SU(2)_L\times SU(2)_R$, and let
\begin{equation}
        S_6^+=\mathbf 4,
        \qquad
        S_6^- =\bar{\mathbf 4}
\end{equation}
denote the two chiral spinor representations of $\Spin(6)\cong SU(4)$, up to
the orientation convention.  The ten-dimensional chirality operator factorizes
as
\begin{equation}
        \Gamma_{11}^{(10)}=\gamma_5^{(4)}\otimes \Gamma_7^{(6)}.
\end{equation}
Therefore a positive ten-dimensional Weyl spinor decomposes as
\begin{equation}
\boxed{
        S_{10}^{+}\rightarrow
        (S_4^L\otimes S_6^+)\oplus(S_4^R\otimes S_6^-)
        }
\label{eq:positive10dbranch}
\end{equation}
or, in representation notation,
\begin{equation}
\boxed{
        \mathbf{16}_{+}^{\rm Lorentz}\rightarrow
        (2,1;\mathbf 4)\oplus(1,2;\bar{\mathbf 4}) .
        }
\label{eq:positive10dbranch-reps}
\end{equation}
Similarly,
\begin{equation}
\boxed{
        S_{10}^{-}\rightarrow
        (S_4^L\otimes S_6^-)\oplus(S_4^R\otimes S_6^+)
        }
\label{eq:negative10dbranch}
\end{equation}
or
\begin{equation}
\boxed{
        \mathbf{16}_{-}^{\rm Lorentz}\rightarrow
        (2,1;\bar{\mathbf 4})\oplus(1,2;\mathbf 4) .
        }
\label{eq:negative10dbranch-reps}
\end{equation}
The dimensions check immediately: each summand has dimension $2\times4=8$, so
a ten-dimensional Weyl spinor has complex dimension $16$.

For the product compactification, the internal calculation gives
\begin{equation}
        \dim\ker\slD_{X_6}^{+}=3,
        \qquad
        \dim\ker\slD_{X_6}^{-}=0
\end{equation}
for $(m,n)=(3,3)$, with the chosen orientation.  If the ten-dimensional fermion
has positive Lorentz chirality, only the first summand in
Eq.~\eqref{eq:positive10dbranch} contributes to massless four-dimensional zero
modes.  Thus the surviving fields are left-handed four-dimensional Weyl
spinors.  Choosing the opposite ten-dimensional chirality, or reversing the
orientation/monopole sign, reverses the four-dimensional chirality.

The full fermion field used below should therefore be written as
\begin{equation}
        \Psi\in \Gamma(S_{10}^{+}\otimes \mathbf{16}_G),
\label{eq:fermion-bundle-precise}
\end{equation}
not simply as ``a $\mathbf{16}$ of $\Spin(10)$.''  With the orientation and
chirality convention above, the three internal zero modes produce
\begin{equation}
\boxed{
        3\times \mathbf{16}_{G,L}
        }
\label{eq:three-left-16s}
\end{equation}
in four dimensions, with no conjugate vectorlike partner coming from the
internal kernel.

Let $\eta_I(y)$, $I=1,2,3$, denote a basis of internal product zero modes for
$(m,n)=(3,3)$.  Expanding the ten-dimensional fermion as
\begin{equation}
        \Psi(x,y)=\sum_{I=1}^3 \psi_I(x)\otimes \eta_I(y)+\cdots,
\end{equation}
where the ellipsis denotes massive Kaluza-Klein modes, gives three massless
four-dimensional chiral fermion multiplets before electroweak symmetry breaking
and other dynamical mass effects.

We take the visible gauge representation for one product zero mode to be the
standard $\mathbf{16}_G$ of $\Spin(10)_G$ reviewed in
Sec.~\ref{sec:history}.  Combining this gauge choice with the Lorentz branching
in Sec.~\ref{subsec:lorentz-branching}, the product zero-mode spectrum is
therefore
\begin{equation}
        3\times \mathbf{16}_{G,L},
\end{equation}
or, after branching to the Standard Model subgroup, three left-handed copies of
Eq.~\eqref{eq:knownspin10decomp}.  Because one $\mathbf{16}_G$ is anomaly-free
as reviewed above, the three-copy spectrum is anomaly-free in four dimensions.

\begin{table}[t]
\caption{Resulting product zero-mode spectrum for $(m,n)=(3,3)$, expressed as
three copies of the standard $\mathbf{16}$ branching.}
\begin{ruledtabular}
\begin{tabular}{ccc}
Field & $SU(3)_C\times SU(2)_L$ & Multiplicity \\
\hline
$Q_L$ & $(3,2)_{1/6}$ & $3$ \\
$u_L^c$ & $(\bar 3,1)_{-2/3}$ & $3$ \\
$d_L^c$ & $(\bar 3,1)_{1/3}$ & $3$ \\
$L_L$ & $(1,2)_{-1/2}$ & $3$ \\
$e_L^c$ & $(1,1)_1$ & $3$ \\
$N_L^c$ & $(1,1)_0$ & $3$ \\
\end{tabular}
\end{ruledtabular}
\end{table}

\section{Gauge-singlet component of the zero-mode spectrum}
\label{sec:singlet}

The spectral calculation also identifies a gauge-singlet component in each
family.  Each $\mathbf{16}_{G,L}$ contains
\begin{equation}
        N_L^c\sim (1,1)_0,
\end{equation}
so the product compactification contains three Standard-Model singlets together
with the three charged families.  Since these states are neutral under
$SU(3)_C\times SU(2)_L\times U(1)_Y$, a Majorana mass term for them is allowed
by the visible gauge symmetry,
\begin{equation}
        \mathcal L_M=-\frac12 M_R N_L^cN_L^c+\mathrm{h.c.}
\end{equation}
The present paper does not compute $M_R$ or any associated neutrino-sector
dynamics.  The only result used here is the group-theoretic statement that the
same index construction producing three charged families also produces three
visible-gauge singlet components.

In an Einstein-Cartan embedding, elimination of the non-propagating torsion
generates a four-fermion contact interaction whose projection onto this
unprotected singlet channel induces a Planck-scale Majorana mass,
$M_R\sim M_{Pl}$.  The derivation of this scale, the role of the
spin-torsion condensate in moduli stabilization, and the identification
of the heavy singlets as candidates for geometric dark matter are
developed in the companion paper \cite{ShayaUECKK2026}.  The present paper
establishes the existence, multiplicity, and overlap structure
(Sec.~\ref{sec:overlap}) of the chiral zero modes on which that
analysis is built.

\section{Zero-mode overlap structure}
\label{sec:overlap}

For applications in which the zero modes enter effective four-dimensional
operators, the relevant geometric data include mode-overlap integrals.  We
collect the basic overlaps here because they are intrinsic properties of the
zero-mode sector constructed above and do not require specifying a particular
low-energy mass-generation mechanism.

\subsection{Explicit basis at $(m,n)=(3,3)$}

The three $S^2$ zero modes are sections of
$S^{+}_{S^2}\otimes\mathcal O(3)\simeq\mathcal O(2)$.  In the
stereographic coordinate $z$ and unitary trivialization,
\begin{equation}
\chi_r(z)=\sqrt{\frac{3}{4\pi a_1^2}}\,\binom{2}{r}^{1/2}
\frac{z^r}{1+|z|^2}\,e_+,\qquad r=0,1,2,
\label{eq:zm-basis}
\end{equation}
with $e_+$ the positive-chirality spinor.  These are orthonormal,
$\int_{S^2}\chi_r^\dagger\chi_s\,\sqrt{g}\,d^2x=\delta_{rs}$.  The
$\mathbb{C}P^2$ factor contributes the single canonical $\Spinc$ mode
$\xi_0=V_{\mathbb{C}P^2}^{-1/2}\,e_{00}$, constant in the unitary
gauge.  The product modes are $\eta_I=\chi_{I-1}\otimes\xi_0$,
$I=1,2,3$.

\subsection{Projector kernel}

The zero-mode projector $P_0(y,y')=\sum_I\eta_I(y)\eta_I^\dagger(y')$
has the coincidence limit
\begin{equation}
P_0(y,y)=\frac{3}{V_6}\,\Pi_0^{\rm int},
\qquad V_6=4\pi a_1^2\,V_{\mathbb{C}P^2},
\label{eq:P0-coincidence}
\end{equation}
with $\Pi_0^{\rm int}$ the rank-one internal spinor projector: the
mode density is exactly homogeneous, as required by the symmetry of
the background.  At separated points the binomial sum collapses,
\begin{align}
\sum_r\binom{2}{r}(z\bar z')^r&=(1+z\bar z')^2,\\
\operatorname{tr}\!\big[P_0(y,y')P_0(y',y)\big]
&=\frac{9}{V_6^2}\cos^4\!\frac{\gamma}{2}.
\label{eq:P0-kernel}
\end{align}
where $\gamma$ is the angular separation on $S^2$---the standard
coherent-state overlap of $\mathbb{C}P^1$.

\subsection{Quartic overlap tensor}

Define
$\mathcal G_{IJKL}=\int_{X_6}dV_6\,
(\eta_I^\dagger\eta_J)(\eta_K^\dagger\eta_L)$.
The azimuthal integral enforces the selection rule
$r_I+r_K=r_J+r_L$, and the radial Beta integral gives, with
$p=r_I+r_K$ and $C_I\equiv\binom{2}{r_I}$,
\begin{equation}
\boxed{\;
\mathcal G_{IJKL}
=\frac{9}{V_6}\,\sqrt{C_IC_JC_KC_L}\;
\frac{p!\,(4-p)!}{5!}\;
\delta_{\,r_I+r_K,\,r_J+r_L}\; }
\label{eq:G-tensor}
\end{equation}
Both independent contractions evaluate to
\begin{align}
\sum_{I,K}\mathcal G_{IIKK}
&=\int_{X_6}dV_6\Big(\sum_I|\eta_I|^2\Big)^2=\frac{9}{V_6},
\label{eq:G-density}
\\
\sum_{I,J}\mathcal G_{IJJI}
&=\int_{X_6}dV_6\,\operatorname{tr}P_0(y,y)^2=\frac{9}{V_6},
\label{eq:G-exchange}
\end{align}
consistent with Eqs.~\eqref{eq:P0-coincidence} and
\eqref{eq:P0-kernel}.  The generalization of the $9$ to
$N_{\rm fam}^2$ for general admissible flux follows from two
properties rather than explicit recomputation.  First, transitivity of
the isometry group ($SU(2)\times SU(3)$ acting on the product) forces
the coincidence density to be homogeneous, so
$\operatorname{tr}P_0(y,y)=N_{\rm fam}/V_6$ identically.  

Second, whenever the index is saturated by sections of a single line bundle, all zero modes multiply the \emph{same} internal spinor line. This occurs for the $S^2$ tower at any $m$, where all modes carry $e_+$, and for the $\mathbb{C}P^2$ factor whenever the kernel is saturated in $\Lambda^{0,0}$, as proven for $n=3$ in Sec.~\ref{sec:nopairs} and for $n=5$ in Sec.~\ref{subsec:one-five}. In that case the coincidence projector is rank one,
$ P_0(y,y)=\frac{N_{\rm fam}}{V_6}\,|\chi\rangle\langle\chi|$,
and both contractions equal $N_{\rm fam}^2/V_6$ exactly.

If saturation fails
and the kernel splits across internal chirality blocks of dimensions
$N_b$, the exchange contraction becomes $\sum_b N_b^2/V_6$, bounded
between $N_{\rm fam}/V_6$ and $N_{\rm fam}^2/V_6$; the quoted
$N_{\rm fam}^2$ scaling therefore assumes saturation, which is the
case proved here for $n=3$ and $n=5$, and more generally for all
odd $n\geq3$ in Sec.~\ref{subsec:general-saturation}.

\subsection{Spin-2 structure}

Under the $SU(2)$ isometry the three modes form a spin-1 triplet, so
the symmetric pair space decomposes as
$\mathrm{Sym}^2(\mathbf 3)=\mathbf 5\oplus\mathbf 1$.  The overlap
tensor is proportional to the spin-2 projector,
\begin{equation}
\mathcal G_{IK;JL}=\frac{9}{5V_6}\,P^{(2)}_{IK;JL}\,,
\label{eq:G-spin2}
\end{equation}
and annihilates the singlet
$(|22\rangle-|13\rangle-|31\rangle)/\sqrt3$, as direct evaluation of
Eq.~\eqref{eq:G-tensor} confirms.  Quartic zero-mode interactions on
this background are therefore pure quadrupole in mode space.

\section{Leading Kaluza-Klein scales in the product model}
\label{sec:kk}

The exact massive spectrum depends on the full background connection, the
choice of metric normalization, and possible torsion.  Nevertheless, the
leading product scale is fixed dimensionally by the inverse internal radii,
\begin{equation}
        M_{\rm KK}\sim \min(a_1^{-1},a_2^{-1}).
\end{equation}
The known $S^2$ monopole tower reviewed in Eq.~\eqref{eq:knowns2spectrum} gives,
for the three-generation choice $m=3$,
\begin{equation}
        M_{1,S^2}=\frac{2}{a_1} .
\end{equation}
For the $\CP^2$ factor, the full nonzero spectrum can be obtained by harmonic
analysis on the homogeneous space $SU(3)/U(2)$ with the chosen $\Spinc$ bundle.
For the purposes of the present product-index paper, the important point is
that the nonzero $\CP^2$ tower is separated from the zero-mode sector by a gap
of order $a_2^{-1}$,
\begin{equation}
        M_{\CP^2,1}=c_{\CP^2}(n)\,a_2^{-1},
\end{equation}
where $c_{\CP^2}(n)$ is a dimensionless number determined by the precise metric
and bundle normalization.  The value of this coefficient controls the first
massive $\CP^2$ excitation.  It is not needed to establish the absence of
additional zero modes: for $n=3$, Sec.~\ref{sec:nopairs} proves by the
Lichnerowicz-Dolbeault argument that the $\CP^2$ kernel consists of exactly one
positive-chirality state and no opposite-chirality partner.

\section{Discussion}

The contribution of this paper is the product use of two known chiral-index
ingredients.  Dolan and Nash established the relevance of $\Spinc$ zero modes
on $\CP^2$ and related complex projective spaces to Standard-Model-like chiral
spectra.  The additional $S^2$ factor introduced here is used not merely as an
extra compact dimension, but as a monopole-controlled replication sector.  The resulting product index is $\ind \slD_{S^2\times \mathbb{C}P^2}=m\,\frac{n^2-1}{8}$.
Two flux sectors give exactly three chiral internal zero modes:
$(m,n)=(3,3)$, where the canonical $\Spinc$ structure on $\CP^2$
supplies a single chiral block and the $S^2$ monopole replicates it
three times, and $(m,n)=(1,5)$, where the higher $\Spinc$ twist on
$\CP^2$ itself provides a three-dimensional kernel and a single $S^2$
mode carries it through.  Saturation---the absence of vectorlike
partners---is proved for both sectors, and indeed for all odd
$n\geq3$, by Kodaira vanishing on the twisted spinor bundle
(Sec.~\ref{subsec:general-saturation}).  The two sectors are
degenerate in family number but differ in their internal wave-function
profiles: in $(3,3)$ the families are distinguished by their $S^2$
angular structure, while in $(1,5)$ they are distinguished by their
$\CP^2$ polynomial degree.  Taking the ten-dimensional fermion to lie
in $S_{10}^{+}\otimes\mathbf{16}_G$ gives, by the Lorentz branching
$\Spin(1,9)\supset\Spin(1,3)\times\Spin(6)$, three left-handed
$\mathbf{16}_{G,L}$ multiplets in either sector: three anomaly-free
Standard Model families plus a right-handed neutrino per family.

Several points are intentionally not claimed here.  The calculation
does not by itself derive the Higgs sector, Yukawa textures, moduli
stabilization, the observed charged-fermion masses, sterile-neutrino
masses, or a dynamical selection principle between $(3,3)$ and $(1,5)$
or among any of the higher-family sectors.  It shows that once a
topological sector is chosen, the generation count is fixed by the
product index and is not adjustable by small smooth deformations of
the metric, torsion, or gauge connection.  Which sector is
cosmologically preferred is a dynamical question addressed in the
companion paper \cite{ShayaUECKK2026}.  The gauge-singlet $N_L^c$
components identified in Sec.~\ref{sec:singlet} are part of the
zero-mode spectrum in every sector, but their masses and
phenomenological roles require additional dynamics beyond the scope
of this index-theoretic paper.

\bibliographystyle{apsrev4-2}
\bibliography{zero_modes}

\end{document}